\documentclass[12pt]{article}
\usepackage{epsfig}
\usepackage{amssymb,graphicx}
\def\be{\begin{equation}}
\def\ee{\end{equation}}
\def\ba{\begin{array}{c}}
\def\ea{\end{array}}

\def\ben{\[}
\def\een{\]}

\newcommand{\bea}{\begin{eqnarray}}
\newcommand{\eea}{\end{eqnarray}}

\begin{document}


\begin{center}

{\Large \bf {

Semi-analytic version of shooting method and the bound states in
non-analytic potential $V(x)= -g^2\exp (-|x|)$

 }}

\vspace{13mm}

 {\bf Miloslav Znojil}

 \vspace{3mm}
Nuclear Physics Institute ASCR, Hlavn\'{\i} 130, 250 68 \v{R}e\v{z},
Czech Republic

 e-mail:
  znojil@ujf.cas.cz

\vspace{3mm}


\end{center}

\subsection*{Keywords:}

quantum bound states; majorization of energies; two-sided
bracketing; Bessel functions; regular solutions; shooting;

.

PACS 03.65.Ge – Solutions of wave equations: bound states

PACS 02.30.Gp – Special functions

%


\section*{Abstract}

People construct quantum bound states either non-numerically (in not
too many exceptional cases) or, in general, numerically (mainly
using majorization or discretization techniques). We argue that
there exists a class of interactions ``in between'' being, in
conventional terminology,  neither exactly solvable nor purely
numerical. The idea is illustrated by the elementary
exponential-well toy-model potential $V(x)= -g^2\exp (-|x|)$.

\newpage

\section{Introduction}

For many smooth and confining phenomenological potentials $V(x)$
which vanish in infinity a quick and, often, fairly reliable upper
estimate of their low-lying bound-state energies $E_0, E_1, \ldots$
is provided by the elementary approximation or majorization of
Hamiltonian using a suitable harmonic-oscillator well
 \be
 V_+^{(HO)}(x)= -M+\omega^2x^2\,.
 \label{major}
 \ee
In practice and, in particular, for the spiked shapes of potentials
$V(x)$ the approximation or majorization of such an oversimplified
type need not work too well of course. For illustration let us
consider bound-state Schr\"{o}dinger equation
 \be
 -\, \frac{{\rm d}^2}{{\rm d} x^2} \psi_n(x)
 + V(x) \psi_n(x)= E_n\,
 \psi_n(x)\,,
 \ \ \ \ \ \
 \psi_n(x) \in L^2(\mathbb{R})\,,\ \ \ \ n = 0, 1, \ldots
   \label{SEx}
  \ee
with one of the most elementary potential wells (cf. \cite{Ryu})
 \be
 V(x)= -g^2\exp (-|x|)\,,
 \ \ \ \ x \in (-\infty,\infty).
 \label{tenpo}
 \ee
This potential is spiked in the origin and decreases, exponentially,
at the large $x$. Once we decide to perform the above-mentioned
majorization we find out that this approach does not lead to
satisfactory results (cf. section \ref{app} below for more details).

Alternatively we may make use of the standard numerical
discretization techniques \cite{Acton,Ros} and we may successfully
guarantee, at any trial energy $E$, the validity of the
above-mentioned normalizability condition $\psi(x) \in
L^2(\mathbb{R})$ by the (in general, numerical, discretization)
construction of the asymptotically correct (often called ``Jost
solutions'' \cite{Newton}) $\psi(x)=\psi_{Jost}(x,E)$. In this
setting the role of the secular equation becomes played by the
condition of the matching of the logarithmic derivatives of the left
and right Jost solution in the origin.

In what follows we intend to reveal and emphasize that for the
non-smooth and polynomially unsolvable but still sufficiently
elementary potentials as sampled by Eq.~(\ref{tenpo}) neither of the
above two approaches looks optimal. Firstly, in the context of
majorization we point out that the presence of the downward-oriented
spike in the origin makes the approximation/majorization of function
$V(x)$ by the exactly solvable model (\ref{major}) fairly bad and
insufficient.

In the second context and in the light of the Bessel-function
solvability of differential equation (\ref{SEx}) + (\ref{tenpo}) on
both half axes one feels tempted to take the advantage and to use,
in a way described in Ref.~\cite{Ryu}, the necessary Jost solutions
in their available, closed analytic form. In practical computations,
unfortunately, it is often difficult to guarantee the sufficient
precision of the bound-state energies as obtained from the
Jost-solution matching-generated secular equation (let us add here
that this discouragement is not universally valid - see, e.g., the
recent remarkable amendment of the Jost-matching construction for
the Coulomb-resembling $V(x) = 1/\sqrt{x}$ in paper \cite{Ishkh1}).

Our present paper is basically motivated by the latter, in general
discouraging numerical experience. In particular, in section
\ref{stat} we recommend the use of an alternative algorithm called,
often, a ``shooting algorithm'' \cite{Acton,Joli}. In its framework
the matching of the Jost solutions in the origin is replaced by the
fine-tuning of the (in general, unphysical) asymptotics of the so
called regular solutions, i.e., of the alternative Bessel-function
solutions which are constructed as regular in the origin.

Our results and conclusions are shortly discussed in the last
section \ref{discussion}. We emphasize there that the proposed
semi-analytic version of the shooting algorithm appears remarkably
user-friendly.

\section{The majorization technique and the limitations of its efficiency \label{app}}

\begin{figure}[h]                    
\begin{center}                         
\epsfig{file=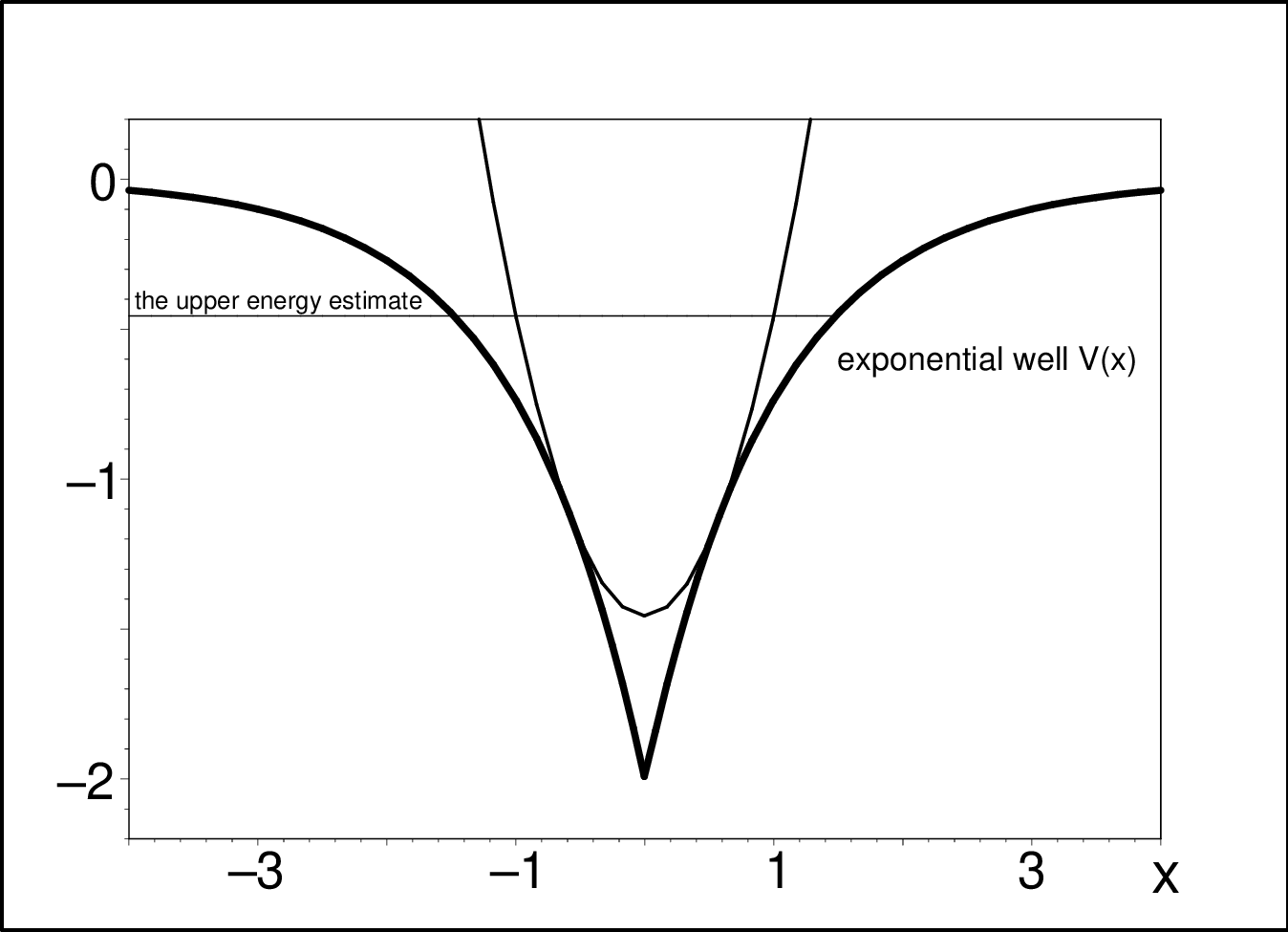,angle=270,width=0.56\textwidth}
\end{center}    
\vspace{2mm} \caption{Sample of majorization of potential
(\ref{tenpo}) with $g = \sqrt{2}$ (thick line) by harmonic
oscillator (\ref{major}) using $\omega=1$ with maximal shift
$M=M(\omega)=1.455938091$ yielding the upper estimate of the
ground-state energy $E_0 < E_+^{(HO)}=-0.455938091$.
 \label{aqizt}
 }
\end{figure}

 \noindent
As long as our benchmark bound-state problem (\ref{SEx}) +
(\ref{tenpo}) does not belong to the traditional exactly solvable
models, it does not look suitable, in spite of its extreme formal
simplicity, to play the role of an illustrative toy model in
introductory textbooks. Moreover, in spite of its interesting
short-range nature resembling. e.g., the popular Yukawa (i.e., a
``short-range Coulomb'') interaction $V^{(Yukawa)}(x) \sim
e^{-x}/x$, we never found the model used in any phenomenologically
oriented tests of approximation techniques. On this background we
felt tempted to apply a few elementary variational approaches to the
upper-bound estimate of ground-state energy $E_0= -k^2$, and we
found the results a bit unexpected.

\subsection{Misbehaved estimates of the ground-state energy
eigenvalues}

Let us now return to our toy model (\ref{SEx}) + (\ref{tenpo}) and
let us consider its most elementary majorization by harmonic
oscillator (\ref{major}). The results are sampled in
Fig.~\ref{aqizt}. Initially, the inspection of the picture made an
impression that the harmonic-oscillator-based upper estimate of the
exact ground state energy $E_0$ cannot be too bad. Unfortunately,
such an expectation appeared not well founded because the precision
of the estimate was found unexpectedly strongly sensitive to our
choice of the freely variable parameters in Eq.~(\ref{major}). In
particular, as long as a routine, brute-force numerical solution of
the problem yields $E_0 \approx -0.817$, we immediately see that the
correct ground-state energy value lies far below the HO-based
upper-bound estimates, especially when obtained at a non-optimal
value of $\omega$.

\begin{figure}[h]                    
\begin{center}                         
\epsfig{file=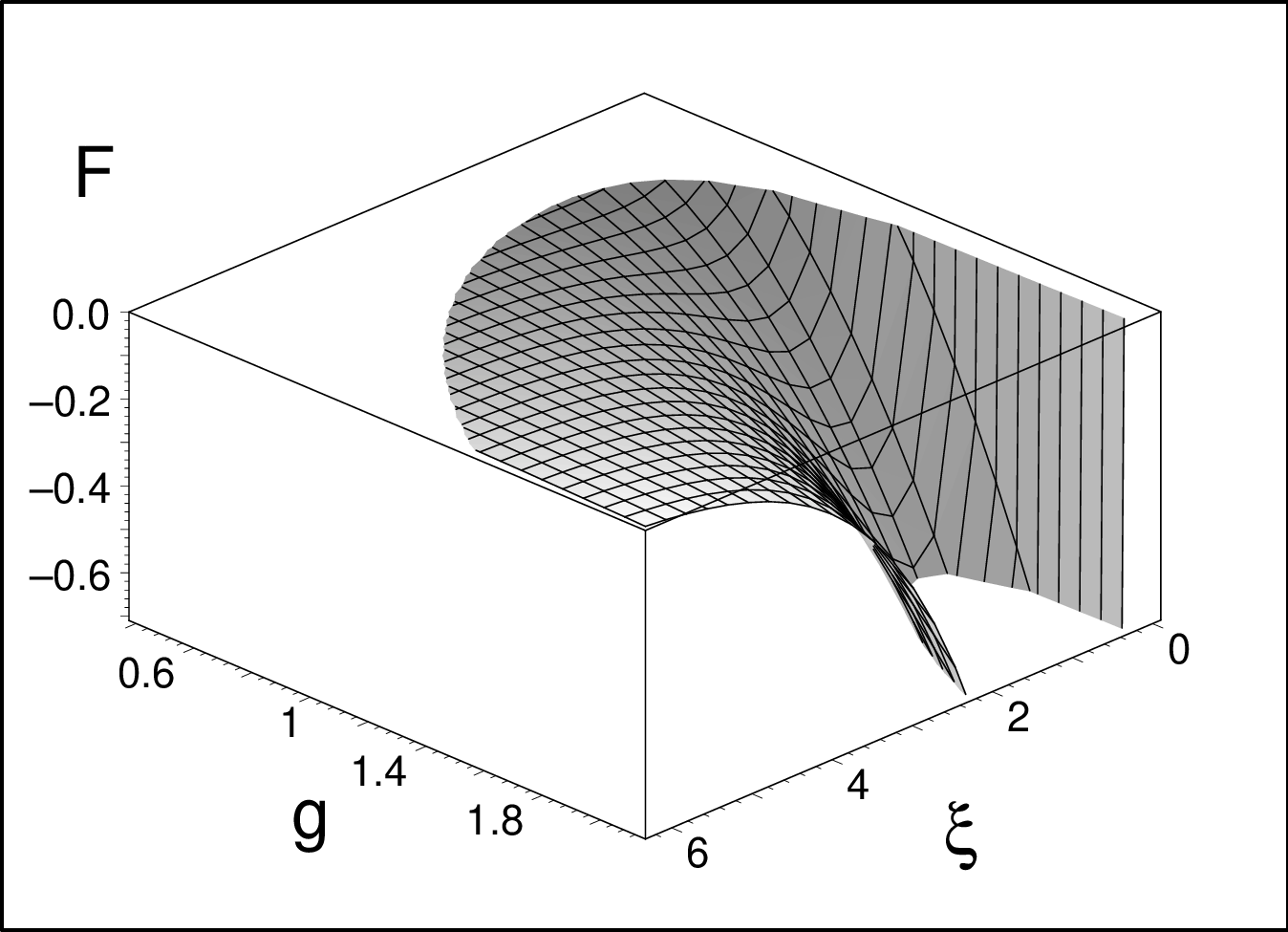,angle=270,width=0.56\textwidth}
\end{center}    
\vspace{2mm} \caption{Upper-bound $F=E_+^{(HO)}$ of the ground state
energy $E_0$ as function of coupling $g$ and of the
osculation-coordinate optional parameter $\xi$.
 \label{svizt}
 }
\end{figure}

The necessity of the careful optimization of the HO estimates is
illustrated in Fig.~\ref{svizt}. The picture offers not only a
strong warning against a naive use of the harmonic-oscillator
majorization but it also illustrates why one should replace the
variationally motivated variability of the spring constant $\omega$
by the variability of the coordinate $x_0$ of the osculation of the
two curves.

A few details of the underlying interesting mathematics are added in
the separate forthcoming subsection - it is slightly more technical
and could be skipped during the first reading.

\subsection{The closed-form construction of the optimal majorizing $V^{(HO)}_+(x)$}

The inspection of Fig.~\ref{aqizt} indicates that once we select the
strength of our one-parametric model [i.e., coupling $g$ in
(\ref{tenpo})] and once we choose an optimal majorizing harmonic
oscillator (i.e., a tentative spring constant $\omega$) we have to
determine the maximal, optimal value of shift $M=M(\omega)$ after
which the two curves [viz., $ V(x)$ of Eq.~(\ref{tenpo}) and
$V_+^{(HO)}(x)$ of Eq.~(\ref{major})] touch at a symmetric pair of
coordinates $x=\pm \xi$. This means that we have to satisfy the pair
of osculation constraints
 \be
 -g^2e^{-\xi}= \omega^2 \xi^2-M\,,
 \ \ \ \ \
 g^2e^{-\xi}= 2\,\omega^2 \xi\,.
 \label{toye}
 \ee
Treating $\xi$ as a new free parameter we have, from the latter
relation,
 \be
 \omega=\omega(\xi)=\frac{g}{\sqrt{2\xi e^\xi}}\,,
 \ee
i.e., a one-to-one correspondence between small $\xi$ and large
$\omega$ and {\it vice versa}. Next, the sum of relations
(\ref{toye}) yields the second necessary definition
 \be
 M=M_{(\xi)}(\omega)=\omega^2(\xi) \,\left (\xi^2+2\,\xi
 \right )
 =g^2\,e^{-\xi}\left (1+\xi/2 \right )
 \,
  \ee
which is a smoothly decreasing function of $\xi$ as well. Thus,
there emerges a conflict between the decrease and increase of the
respective components of the resulting $\xi-$dependent upper bound
 \be
 E_n^{(HO)}=E_n^{(HO)}(g,\xi)=(2n+1)\,\omega(\xi)-M
 =\frac{ge^{-\xi/2}}{\sqrt{2\xi} }(2n+1)
 -g^2\,e^{-\xi}\left (1+\xi/2 \right )\,.
 \label{looksc}
 \ee
At $n=0$ the shape of this function is sampled in Fig.~\ref{svizt}
showing that at every given coupling $g$ there exists a sharp
minimum of the bound $E_n^{(HO)}(g,\xi)$ at a unique optimal value
of $\xi=\xi_0(g)$.

Formula (\ref{looksc}) for function $E_n^{(HO)}(g,\xi)$ of the two
variables looks complicated but we may still try to determine the
curve $\xi_0(g)$ of the minima at a fixed coupling. The problem
looks complicated because the corresponding necessary condition
$\partial_\xi E_n^{(HO)}(g,\xi_0)=0$ leads to a complicated equation
for $\xi_0(g)$. Fortunately, a pleasant surprise emerges after one
employs a computer-assisted symbolic manipulation solver and obtains
the inverse function $g(\xi_0)$ in an unexpectedly elementary form
 \be
 g(\xi_0)=\sqrt{{\frac {{}{e^{\xi_0}}}{2{\xi_0}^{3}}}}\,.
 \label{thecur}
 \ee
In a graphical representation this is a smooth U-shaped curve with a
unique minimum at $\xi_0=3$. Its shape is determined by the decrease
$\sim \xi_0^{-3/2}$ at small $\xi_0$ and by the increase $\sim
\exp(\xi_0/2)$ at large $\xi_0$.

We would like to add that the value of $\xi_{critical} =3$ separates
the ``curve of extremes'' (\ref{thecur}) into its left, $\xi_0
=\xi_0^{(useful)} \in (0, 3)$ branch (carrying the minima {\it
alias} the optimal estimate values of $E_n^{(HO)}(g(\xi_0),\xi_0)$)
and the spurious branch of the variationally useless (and, moreover,
even positive, physically useless) maxima of the HO-based  upper
bound at $\xi_0=\xi_0^{(useless)} \in (3,\infty)$.

\section{Semi-analytic implementation of the shooting method\label{stat}}

As we already mentioned, people often construct, numerically, the
two asymptotically correct branches of wave functions and match
their logarithmic derivatives near the minimum of $V(x)$ (cf.
subsection \ref{para}). Using a toy model $V(x)= -g^2\exp (-|x|)$ we
shall now discuss the situation in which such an approach leads to
the less satisfactory results. An optimal strategy of the
determination of the spectra will be then recommended as provided by
the shooting algorithm which is initiated in the origin.

\subsection{The Jost-solution technique and the limitations of its
efficiency\label{para}}

Let us first return to the search for an effective alternative to
the above-criticized majorization approximation approach. Let us
also recollect that for potentials sampled by Eq.~(\ref{tenpo}) it
is really possible to search for the bound-state solutions via a
proper matching of the left and right branches of the well known and
asymptotically correctly vanishing Bessel-function solutions $\sim
Z_\nu(y)$ with suitable index $\nu=\nu(E_n)$ and argument $y=y(x)$.
Finally, let us remind the readers that the practical implementation
and realization of such a form of matching in the origin may
encounter numerous, mostly purely numerical obstacles.

One of them is connected with the common strategy of the
trial-and-error choice and systematic amendment of the variable
candidate-for-the-energy parameter $E=E_{(trial)}$ in
Eq.~(\ref{SEx}). Whenever one knows the general solution of the
ordinary differential Schr\"{o}dinger equation in the form of
superposition $\psi(x)=c_1 \psi_1(x)+c_2\psi_2(x)$ of the two known
components $\psi_{1,2}(x)$, and whenever one has to match them in
the origin (which is also one of the promising options for our toy
model (\ref{tenpo}) where the available wave function components are
Bessel functions), one usually starts from imposing the correct
asymptotic behavior analytically. Then one completes the
construction via the ``smoothness-in-the-origin'' strategy, i.e.,
via the matching of the logarithmic derivatives of the candidate
wave functions at $x=0$.

In the algorithm, the main theoretical difficulty may be seen in the
lack of a friendly representation of the asymptotically correct
Bessel wave functions $\sim Z_\nu(y)$ at small $y$ and near the
origin \cite{Ryshik}. Another, more practical difficulty reflects
the smallness of changes of the resulting left and right $x \to
0^\pm$ limits of $\psi(x)$ after a small change of the value of
$E_{(trial)}$. In this sense, irrespectively of the representation
of wave functions, the localization of the correct bound-state
energy may prove slow and marred by numerical errors. This
observation inspired also our present alternative methodical
proposal of an opposite strategy.

\subsection{The merits of the analytic shooting method}

We shall now change the strategy and start from a guarantee of the
correct behavior of the superpositions of Bessel-function solutions
$\psi(x)=c_1 \psi_1(x)+c_2\psi_2(x)$ in the origin. We will be able
to take advantage not only of having the general solution
represented by Bessel functions $Z_{\nu(E)}[y(x)]$ but also of the
more easily satisfied duty of the localization of the correct
bound-state energy $E_n(g)$ via the localization of the largest zero
of $\psi(x)$.

Naturally, due to the left-right symmetry of potential (\ref{tenpo})
one has to distinguish between the even-parity states (with property
$\psi(x)=\psi(-x)$, i.e., $\psi(0)=1$ and $\psi'(0)=0$) and the
odd-oparity states (with property $\psi(x)=-\psi(-x)$, i.e.,
$\psi(0)=0$ and $\psi'(0)=1$). Without any significant loss of
insight in the problem let us consider, for the sake of brevity,
just the former case involving the ground state with $n=0$ and all
of the excited states with even quantum numbers $n=2m$.

We may stay, without loss of generality, on the positive half-axis
of $x>0$. Via an elementary change of variables in Eqs.~(\ref{SEx})
+ (\ref{tenpo}) the exact solvability of the model proves equivalent
to the existence of all of the general even-parity wave functions in
the non-numerical form
 \be
 \psi\left( x \right) =
 D_1(k,g)\,{\it {}J}_{-2\,k} \left( 2\,g\,{e^{
 -1/2\,x}} \right) +
 D_2(k,g)\,{\it {}Y}_{-2\,k} \left( 2\,g\,{e^{-1/2\,x}}
 \right)\,.
 \label{geneso}
 \ee
Here we may evaluate, after slightly tedious calculations,
 \ben
 D_1(k,g) =
 -{\frac {  {\it {}Y}_{-2\,
 k+1} \left( 2\,g\, \right) +k\,{\it {}Y}_{-2\,k} \left( 2\, g
  \right)/g    }{{\it {}J}_{-2\,k+1} \left( 2\,g\,
 \right) {\it {}Y}_{-2\,k} \left( 2\,g\, \right) -{\it {}Y}_{-2\,k+1}
 \left( 2\,g\, \right) {\it {}J}_{-2\,k} \left( 2\,g  \right) }}
  \een
  and
  \ben
  D_2(k,g) ={\frac {  {\it {}J}_{-2\,
 k+1} \left( 2\,g\, \right) +k\,{\it {}J}_{-2\,k} \left( 2\, g
  \right)/g    }{{\it {}J}_{-2\,k+1} \left( 2\,g\,
 \right) {\it {}Y}_{-2\,k} \left( 2\,g\, \right) -{\it {}Y}_{-2\,k+1}
 \left( 2\,g\, \right) {\it {}J}_{-2\,k} \left( 2\,g  \right)
 }}\,.
 \een
In the subsequent step we simply choose a tentative value of $k$,
plot the function (\ref{geneso})  of variable $x \in (0, \infty)$
and count the number ${\cal Z}$ of its zeros.

In the search for the exact bound-state energy $E_{2m}(g)$ we have
two possibilities. In the first case marked by (a) with ${\cal Z}<m$
we declare the preselected value of $k$ ``too large''. Having
recalled the standard oscillation theorems \cite{Ince} we choose a
slightly smaller next tentative value of $k$ in order to make our
next, amended candidate (\ref{geneso}) for wave function ``more
oscillatory''. In the second case marked by (b) with ${\cal Z}\geq
m$ we do the opposite. On the basis of the same oscillation theorems
we choose a slightly larger (but not ``too large'') next tentative
value of $k$ in order to make our amended candidate (\ref{geneso})
for wave function ``less oscillatory''.

In practice, the graphical determination of the actual nodal count
${\cal Z}$ proves easily implemented because if the maximal node
disappears in infinity, the asymptotic, exponentially quickly
growing part of the tentative regular solution $\psi(x)$ must change
its sign. This makes the resulting ``analytic shooting method''
fully analogous to its universal, quickly convergent numerical
predecessors \cite{Newton}.

%
%
%

\begin{figure}[h]                    
\begin{center}                         
\epsfig{file=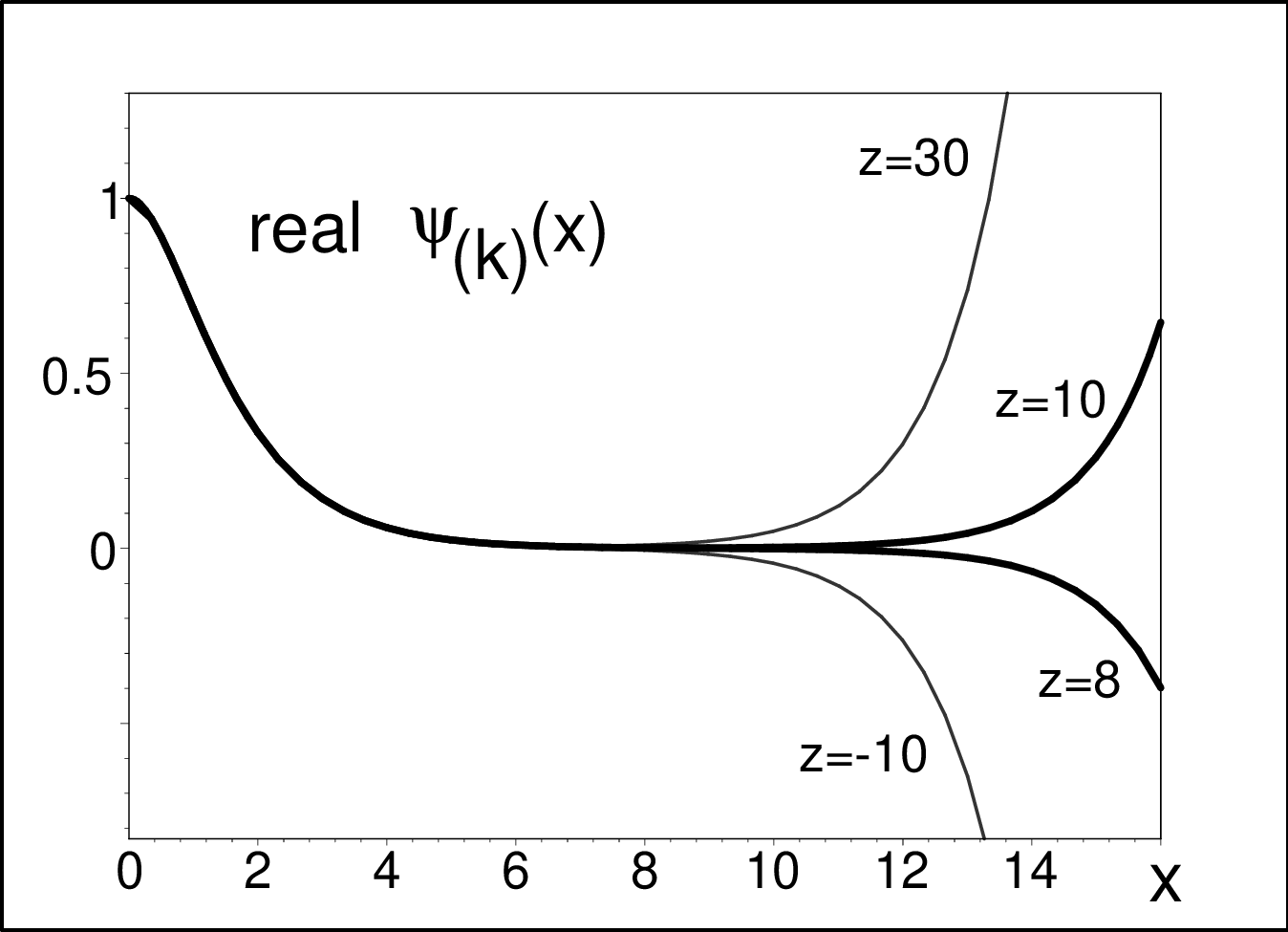,angle=270,width=0.56\textwidth}
\end{center}    
\vspace{2mm} \caption{The sample of the changes of the asymptotics
of the regular and real $x>0$ part of the ground-state wave function
with the variation of the error term $z$ in $k=k(z)$ at the
even-parity normalization $\psi_{(k)}(0)=1$,  $\psi'_{(k)}(0)=0$ and
at the fixed coupling $g=\sqrt{2}$.
 \label{soizt}
 }
\end{figure}

Typically, the implementation of the analytic shooting algorithm is
easy as it may proceed simply by halving the interval of eligible
$k$s. Another, serendipitous advantage of the algorithm is its
connection with oscillation theorems so that it generates, in
practice, the sequence of parallel upper and lower bounds of the
energies. For example, the above-mentioned approximate ground-state
result $E_0=-k_0^2\approx -0.817$ becomes replaced, in this manner,
by the two-sided estimate $-0.81721 < E_0< -0.81720$ and,
subsequently, by the series of its higher-precision amendments
whenever needed. For an illustration of the mechanism of this
construction let us choose again $g=\sqrt{2}$ and make an ansatz
$E_0=-k_0^2=-0.817200 - z\,10^{-6}$ containing a not yet specified
error term $z$. Naturally, the ``shooting'' choice of the
trial-and-error value of $z$ controls the behavior of the maximal
nodal zero of the related regular solution $\psi(x)=\psi_{(k)}(x)$.
For the four alternative choices of $z$ this behavior is displayed
in Fig.~\ref{soizt}.

\section{Discussion\label{discussion}}

We may conclude that the model $V(x)= -g^2\exp (-|x|)$ exemplifies,
nicely, the situation in which several standard approaches do not
lead to a satisfactory result and in which an optimal strategy of
the determination of the spectra is found in the shooting method
initiated in the origin.

The unlimited-precision nature of the resulting bound-state energies
is in one-to-one correspondence with the unlimited-precision
availability of the Bessel-function wave-function candidates. In
this sense one encounters a big difference between the models
sampled by Eq.~(\ref{tenpo}) (for which we would endorse the ``exact
solvability'' status, cf. also \cite{Ryu}) and the more general
interaction potentials for which the Jost or regular solutions must
be generated by an approximate variational or discretization method.

This seems to imply that the status of exactly solvable (ES)
one-dimensional quantum potentials $V(x)$ should be attributed not
only to the conventional ES family (yielding wave functions
proportional to classical orthogonal polynomials) and not only to
the non-analytic square wells and point interactions (solvable by
the matching techniques) but also to some interactions ``in
between''.

The most characteristic feature of the $k-$dependent candidates for
the wave functions is that these functions remain proportional to
special functions. Near each physical bound-state value of parameter
$k$ they remain normalizable and have to be matched in the origin of
{\it vice versa}. Of course, this is a generic idea which might be
applied to a fairly wide class of confining potentials. Needless to
add that in our present model its implementation was perceivably
facilitated by the parity-symmetry of the model.

Naturally, the theoretical roots of the whole idea are already very
well known. Their origin dates back to the studies of the singular
Sturm-Liouville problems as pioneered by H. Weyl in 1909 and as
explained, e.g., in Chapter 9 of the classical
monograph~\cite{Coddington}. In the language of mathematicians one
could change the conventions and speak also, equivalently, about the
study of the Titchmarsh-Weyl $m$-functions with the property that
their poles correspond to the bound-state eigenvalues, etc.

In the opposite, fully practical context of quantum mechanics of
simple atomic and molecular systems, traditionally, people make
distinction between the general, numerical, and special, exactly
solvable (ES) bound-state Schr\"{o}dinger equation. In this context
Eq.~(\ref{SEx})  is usually assigned the SE status just for the
square-well real and confining $V(x)$ \cite{Constantinescu} (or,
{\it in extremis}, to the non-analytic point-interaction models) or,
alternatively, for the strictly analytic potentials $V(x)$ for which
some \cite{Ushveridze} or all \cite{Cooper} of the bound-state wave
functions $\psi_n(x)$ happen to be proportional to polynomials.

Summarizing, by far the most important condition of the efficiency
of the present semi-analytic version of the well known shooting
algorithm lies in the explicit special-function form of the Jost or
regular solutions. Without this condition one would return either
just to the universal and well-known, purely numerical techniques or
to some other, less universal and less known techniques involving,
{\it pars pro toto}, the so called Hill-determinant methods
\cite{HD}, the so called quasi-exact solvability \cite{Coul} etc. In
our present paper we simply promoted, on this basis, the possibility
of having a new, ``third'' class of certain semi-numerical ES (SNES)
potentials $V^{(SNES)}(x)$.

This class was exemplified by the toy-model interaction
(\ref{tenpo}). We demonstrated that our {\em non-analytic}
illustrative SNES model was tractable by the constructive use of the
general {\em regular} solutions of the differential Schr\"{o}dinger
equation which appeared to be tractable, user-friendly and available
in the form of specific, next to elementary special functions, i.e.,
in our case, of Bessel functions.

\subsection*{Acknowledgements}

The work on the project was supported by the Institutional Research
Plan RVO61389005 and by GA\v{C}R Grant Nr. 16-22945S.

\newpage

\end{document}